\documentclass[epj]{svjour}
\RequirePackage{float}
\usepackage{epstopdf}
\usepackage{hyperref}
\usepackage{amssymb}
\usepackage{mathtools}
\usepackage{capt-of}
\usepackage{graphicx}  
\usepackage{dcolumn}   
\usepackage{bm}%

\begin{document}
\title{Inflation in $f(R,T)$ Gravity}
\author{Snehasish Bhattacharjee\inst{1}, J.R.L. Santos\inst{2}, P.H.R.S. Moraes\inst{3} \and P.K. Sahoo\inst{4}
}                     
%
%
\institute{Department of Astronomy, Osmania University, Hyderabad-500007,
India,  Email: snehasish.bhattacharjee.666@gmail.com
\and UFCG - Universidade Federal de Campina Grande - Unidade Acad\^{e}mica de F\'isica,  58429-900 Campina Grande, PB, Brazil, Email: joaorafael@df.ufcg.edu.br
\and Universidade de S\~ao Paulo, Instituto de Astronomia, Geof\'isica e Ci\^encias Atmosf\'ericas, 05508-090 S\~ao Paulo, SP, Brazil, Email: moraes.phrs@gmail.com
\and Department of Mathematics, Birla Institute of
Technology and Science-Pilani, Hyderabad Campus, Hyderabad-500078,
India,  Email:  pksahoo@hyderabad.bits-pilani.ac.in
}
\date{Received: / Accepted: }
%
\abstract{
The article presents modeling of inflationary scenarios for the first time in the $f(R,T)$ theory of gravity. We assume the $f(R,T)$ functional from to be $R + \eta T$, where $R$ denotes the  Ricci scalar, $T$ the trace of the energy-momentum tensor and $\eta$ the model parameter (constant). We first investigated an inflationary scenario where the inflation is driven purely due to geometric effects outside of GR. We found the inflation observables to be independent of the number of e-foldings in this setup. The computed value of the spectral index is consistent with latest Planck 2018 dataset while the scalar to tensor ratio is a bit higher. We then proceeded to analyze the behavior of an inflation driven by $f(R,T)$ gravity coupled with a real scalar field. By taking the slow-roll approximation, we generated interesting scenarios where a Klein Gordon potential leads to observationally consistent inflation observables. Our results makes it clear-cut that in addition to the Ricci scalar and scalar fields, the trace of energy momentum tensor also play a major role in driving inflationary scenarios.
} 
\titlerunning{Inflation in $f(R,T)$ Gravity} 
\authorrunning{S. Bhattacharjee et al.}
\maketitle
\section{Introduction}
\label{sec01}

Inflation is the standard description of the early Universe and has been successful in resolving major cosmological issues, like the flatness, horizon and fine tuning problems  \cite{guth,als,linde}. The formation of large scale structures and anisotropies observed in CMB can be elegantly described in a cosmological scenario where the pre-inflationary perturbations grew super-exponentially over a brief period of time \cite{mukh,guthpi,hawking,star}. 

In the inflationary cosmology, a scalar field called ``inflaton'' has been presumed to give rise to an inflationary era in the framework of General Relativity \cite{guth}. Various types of potentials capable of producing an inflationary scenario have been rigorously investigated and constrained with observational data \cite{hossain,martin1,martin2,geng,huang}.

Departing from scalar field standard models of inflation \cite{guth,als,linde}, it is also possible to approach this scenario via alternative theories of gravity. Inflation has been widely studied in $f(R)$ gravity \cite{sdvk,vk,kkvk,nojiri}, in $f(\mathcal{T})$ gravity \cite{ft}, in $f(\mathcal{G})$ gravity \cite{yi}, for which $R$, $\mathcal{T}$ and $\mathcal{G}$ are the Ricci, torsion and Gauss-Bonnet scalars, respectively, and in braneworld cosmology \cite{Sasaki}. However, as far as the present authors know, inflation has not been addressed in the $f(R,T)$ theory of gravity \cite{harko} this far, for which $T$ represents the trace of the energy-momentum tensor. We shall fill in that gap in this paper. 

The $f(R,T)$ gravity has already provided remarkable results in different areas, such as dark energy \cite{sun/2016}, dark matter \cite{zaregonbadi/2016}, massive pulsars \cite{scmlm/2019,mam/2016}, super-Chandrasekhar white dwarfs \cite{clmaomm/2017}, wormholes \cite{ms/2019,elizalde/2019,mpc/2019,elizalde/2018,ms/2018,sms/2018,smsr/2018,ms/2017,mcl/2017,azizi/2013,sahoo/2020}, bouncing cosmology \cite{bhatta,bhatta2}, big-rip singulairty \cite{rd18}, baryogenesis \cite{bhatta3}, redshift drift \cite{bhatta4}, density fluctuations \cite{bhatta5}, temporally varying physical constants \cite{bhatta6}, Big bang nucleosynthesis \cite{bhatta7} and gravitational waves \cite{sharif/2019,amam/2016}.

The standard method to ascertain the inflation observables is by performing a detailed perturbation analysis \cite{david}. Nonetheless, the approach can be made simpler by introducing the slow-roll parameters \cite{martin2}, either when inflation arises due to gravitational modification or through scalar fields.

Several constraints were imposed on inflation following the release of latest Planck results \cite{planck}. Many inflationary models were excluded due to their inconsistency with observations. Some of the remaining viable models include the Starobinsky model \cite{f13,f14}, the Higgs model \cite{f15} and $\alpha$-attractors \cite{f16,f17,f18}. 

The present paper is organized as follows: In Section \ref{sec02} we provide a summary of $f(R,T)$ gravity. In Section \ref{sec03} we introduce the concept of modeling an inflationary scenario and obtain the expressions of the inflation observables for our proposed model. We then proceed to confront the viability of our model against various cosmological observations. Section \ref{sec04} presents a scenario where the $f(R,T)$ gravity is coupled with an inflaton field. In Section \ref{sec06} we present our conclusions and perspectives.

\section{Overview of $f(R,T)$ gravity}
\label{sec02}

The $f(R,T)$ modified gravity action is given by \cite{harko} 

\begin{equation}
\mathcal{S}=\frac{1}{2\kappa^{2}}\int \sqrt{-g}\left[ f(R,T)+\mathcal{L}_{m}\right] d^{4}x,   
\end{equation}
where $\kappa^{2}=\frac{8 \pi G}{c^{4}}$, $G$ is the Newtonian gravitational constant, $c$ is the speed of light, $g$ is the metric determinant, $\mathcal{L}_m$ is the matter Lagrangian density, which, following \cite{harko}, we take as $-p$, with $p$ being the pressure.

Varying (1) with respect to the metric $g_{\mu\nu}$ generates the $f(R,T)$ gravity field equation as

\begin{equation}
\Sigma_{\mu\nu}f^{1}_{,R}(R,T)+f^{1}_{,R}(R,T)R_{\mu\nu} -\frac{1}{2}g_{\mu\nu}f(R,T)= \kappa^{2}T_{\mu\nu}-f^{1}_{,T}(R,T)(T_{\mu\nu} + \Gamma_{\mu\nu}),
\end{equation}
where $T_{\mu\nu}$ is the energy-momentum tensor, that for a perfect fluid reads 
\begin{equation}
- pg_{\mu\nu} + ( \rho + p)u_{\mu}u_{\nu}= T_{\mu\nu},
\end{equation}
$\rho$ represents the density of the fluid, $u_\mu$ the four-velocity,
\begin{equation}
\Sigma_{\mu\nu}= g_{\mu\nu}\square-\nabla_{\mu} \nabla_{\nu},
\end{equation}
\begin{equation}
\Gamma_{ij}\equiv g^{\mu \nu}\frac{\delta T_{\mu \nu}}{\delta g^{ij}}
\end{equation}
and the notation $f^{i}_{,X}\equiv \frac{d^{i}f}{d X^{i}}$ was implemented.

Now we assume $f(R,T) = R + \eta T$, with constant $\eta$. This is the simplest and most popular $f(R,T)$ gravity model and its viability has been minutely investigated, for instance, in \cite{harko,mam/2016,clmaomm/2017,mpc/2019,ms/2017,mcl/2017,azizi/2013} (check also \cite{m23,m25,m31}). For a flat Friedman-Lem\^aitre-Robertson-Walker space-time with ($-$,$+$,$+$,$+$) metric signature, the modified Friedman equations are given by

\begin{equation}\label{8_1}
H^{2}= \frac{1}{3}\left[-\frac{\eta}{2}\left(\omega - 3\right) + \kappa^{2} \right] \rho,
\end{equation} 
\begin{equation}\label{8}
-3H^{2}-2\dot{H} = \left[-\frac{\eta}{2}\left(1 -3 \omega\right) + \kappa^{2} \omega  \right] \rho,
\end{equation}
in which $\omega = p/ \rho$ is the equation of state parameter and dots represent time derivatives.

Solving the differential equation \eqref{8}, yields

\begin{equation}\label{9}
H=\frac{\sigma}{t},
\end{equation}
with

\begin{equation}\label{10}
\sigma = -\frac{1}{3}\left[ \frac{\left( \omega - 3  \right)\eta - 2\kappa^{2} }{\left(1 + \omega\right)\left(\kappa^{2}  + \eta\right)   }\right] .
\end{equation}

\section{Modeling an inflationary scenario}
\label{sec03}

The primary objective of any theory of gravity to explain inflation successfully is to yield observationally consistent estimate of various inflation-related observables. These are: \textbf{(a)} the scalar spectral index of the curvature perturbations $n_{s}$, \textbf{(b)} the tensor-to scalar ratio $r$ and \textbf{(c)} the tensor spectral index $n_{T}$ \cite{david}. These quantities are constrained very accurately with observational data.

These observables are approximated utilizing slow-roll assumptions, where all the information related to any inflationary scenario is ciphered in the slow-roll parameters. Particularly, one may introduce \cite{martin2}

\begin{equation}
\epsilon_{n+1}=\frac{d}{d N}\log \mid \epsilon_{n}\mid,
\end{equation}
where $N \equiv \log (a/a_{i})$ represents the number of e-foldings, $\epsilon_{0}\equiv H_{i}/H$, where $H_{i}$ and $a_{i}$ represent the Hubble parameter and scale factor at the beginning of inflation, respectively, and $n>0$. Inflation concludes at a scale factor $a_{f}$ when $\epsilon_{1} (a_{f}) \sim 1$ and the slow-roll approximations becomes obsolete. The first two terms of $\epsilon_{n}$ read

\begin{equation}\label{11_1}
\epsilon_{1} = -\frac{\dot{H}}{H^{2}},
\end{equation}  
\begin{equation} \label{11_2}
\epsilon_{2} = -\frac{2 \dot{H}}{H^{2}} + \frac{\ddot{H}}{H \dot{H}}
\end{equation}
and finally the inflation observables ascertained at $a_{i}$ are given by \cite{martin2}

\begin{equation} \label{11_3}
r \approx 16 \epsilon_{1},
\end{equation}
\begin{equation}\label{11_4}
n_{s} \approx 1 - 2(\epsilon_{1}+\epsilon_{2}),
\end{equation}
\begin{equation}\label{11_5}
n_{T}\approx -2 \epsilon_{1}.
\end{equation}

By employing \eqref{9}, we obtain the slow-roll parameters as 

\begin{equation}\label{13}
\epsilon_{1}=-\frac{1}{\sigma}
\end{equation}
\begin{equation}\label{14}
\epsilon_{2}=0
\end{equation}

We assume $\eta = \alpha \kappa^{2}$ where $\alpha$ is a constant and $\omega = -1.028$ (Eq. (50) from \cite{planck}). Therefore  $\sigma$ reads
\begin{equation}\label{15}
\sigma (\omega = -1.028,\eta = \alpha \kappa^{2} ) = \frac{-1}{3}\left[\frac{2 + 4.028 \alpha}{0.028 (1 + \alpha)} \right] 
\end{equation}
Plugging \eqref{15} and \eqref{13} into \eqref{11_3}, \ref{11_4} and \ref{11_5}, we obtain the inflation observables in our $f(R,T)$ gravity model as
\begin{equation}\label{100}
 r = 0.333664 + \frac{0.167992}{0.496524+ \alpha} 
\end{equation}
\begin{equation}\label{99}
n_{s} = 0.958292 - \frac{0.020999}{0.496524+ \alpha} 
\end{equation}
\begin{equation}\label{98}
n_{T} =\frac{2 (1 + \alpha)}{-23.8095 - 47.9524 \alpha}
\end{equation}
We shall now confront the viability of our $f(R,T)$ gravity model with cosmological observations. Planck satellite data constrained the spectral index ($n_{s}$) and scalar to tensor ratio ($r$) at 95\% confidence level as \cite{planck}
\begin{align}
n_{s} = 0.9644 \pm 0.0049,  \hspace{0.5in} r<0.10
\end{align}

\captionof{table}{Inflation observables for different values of $\alpha$
}\label{tab:table}
\begingroup
\setlength{\tabcolsep}{8pt} 
\renewcommand{\arraystretch}{1.5} 
\begin{tabular}{ |p{3cm}|p{3cm}|p{3cm}| p{3cm}| }
 \hline

  $\alpha$   & $r$     & $n_{s}$ & $n_{T}$ \\
    \hline
    0.0 & 0.672  & 0.916  &-0.084  \\
    -1.0 & 0.0  & 1.0  &0.0  \\
     -2.5 & 0.249814  & 0.968773  &-0.0312268  \\
   
    -2.6 & 0.2538  & 0.968275  &-0.031725  \\
    
     -2.7 & 0.257425  & 0.967822  &-0.0321781  \\
     -2.8   & 0.260735 & 0.967408  &-0.0325918\\  
    -2.9    & 0.263769 & 0.967029  & -0.0329711\\  
    -3.0    & 0.266561 & 0.96668  &-0.0333201\\ 
    
 \hline
 
\end{tabular}
\vspace{0.5cm}
\endgroup

Currently, instruments are not sensitive enough to determine $n_{T}$ but with more sensitive devices we ought to be able to constrain it in the near future  \cite{ft,simard}. Though the theoretical estimate of the spectral index $n_{s}$ is well within the observational range, the ascertained estimate of scalar to tensor ratio $r$ is a bit higher. All these observables are extremely sensitive to $\alpha$. For $\alpha = -1$, we obtain $n_{T} = r = 0$ and $n_{s} = 1$, implying the non-existence of any type of inflation in the early Universe.

\section{Inflaton Field}
\label{sec04}

In this section we are going to analyze the behavior of the $f(R,T)$ theory in the presence of a scalar inflaton field. In order to proceed, let us work with the action 

\begin{equation}
S= \frac{1}{2\,k^{\,2}} \int d^4{x}\sqrt{-g}\,\left[f(R,T)+2\,k^{\,2}{\cal L}(\phi,\partial_\mu\,\phi)\right]\,,
\end{equation}
where ${\cal L}$ is a standard Lagrangian density whose form is given by
\begin{equation}
{\cal L}= -\frac{1}{2}\,\partial_{\,\mu}\phi\partial^{\,\mu}\phi-V(\phi)\,,
\end{equation}
and we are assuming
\begin{equation}
\phi=\phi(t)\,; \qquad f=R+\eta\,T\,.
\end{equation}
The components of the energy momentum tensor are derived from the relation
\begin{equation}
T_{\,\mu\,\nu}= g_{\,\mu\,\nu}\,{\cal L}-2\,\frac{\partial\,{\cal L}}{\partial\,g^{\,\mu\,\nu}}\,,
\end{equation}
and their explicit forms are
\begin{equation}
\rho = \frac{\dot{\phi}}{2}+V\,; \qquad p=\frac{\dot{\phi}}{2}-V\,.
\end{equation}

Taking the previous components into the Friedmann Eqs. \eqref{8_1}, and \eqref{8} yields to
\begin{equation}\label{16}
H^2= \frac{1}{6}\,\left[\left(\eta+k^2\right)\,\dot{\phi}+\left(4\,\eta+2\,k^{2}\right)\,V\right]
\end{equation} 
\begin{equation}\label{16_1}
\dot{H}=-\frac{1}{2}\,\left(\eta+k^2\right)\dot{\phi}^{\,2}\,.
\end{equation}
Moreover, the inflaton field $\phi$ must satisfy the equation of motion
\begin{equation} \label{17}
\ddot{\phi}\,\left(\eta+k^2\right)+3\,\left(\eta+k^2\right)\,H\,\dot{\phi}+\left(2\,\eta+k^2\right)\,V_{\phi}=0\,; \qquad V_\phi=\frac{d\,V}{d\,\phi}\,.
\end{equation}

In order to seek for a simple inflationary model, let us use the well-known slow-roll approximation, which states a dynamic regime where
\begin{equation}
\dot{\phi}^{2}\ll V\,; \qquad \ddot{\phi}\ll\dot{\phi}\,,
\end{equation}
consequently, Eqs. \eqref{16} and \eqref{17} are rewritten as
\begin{equation}\label{18}
H^{2}=\frac{1}{3}\,\left(2\,\eta+k^{2}\right)\,V\,; \qquad H\,\dot{\phi}=-\frac{2\,\eta+k^2}{3(\eta+k^2)}\,V_{\,\phi}\,,
\end{equation}
besides,
\begin{equation}
\omega=\frac{\dot{\phi}^2-2\,V}{\dot{\phi}^2+2\,V}\,; \qquad \rightarrow \qquad \omega \approx -1\,,
\end{equation}
in the slow-roll regime. 

Let us consider that 
\begin{equation} \label{18_1}
V=\frac{1}{2}\,m^2\,\phi^2\,,
\end{equation}
meaning that the inflaton rolling is due a Klein-Gordon potential, with $m$ is a mass parameter. Such an assumption yields to
\begin{equation} \label{18_2}
H=\sqrt{\frac{2\,\eta+k^2}{6}}\,m\,\phi\,,
\end{equation}
and
\begin{equation}
\dot{\phi}=-\frac{\sqrt{6\,(2\,\eta+k^2)}}{3(\eta+k^2)}\,m\,.
\end{equation}
This last equation can be directly integrated leading to
\begin{equation}
\phi(t)= \phi_i-\frac{\sqrt{6\,(2\,\alpha+1)}}{3(\alpha+1)}\,m\,t\,,
\end{equation}
where we worked with $\eta=\alpha\,k^2$. Consequently, the previous result means that
\begin{equation}\label{19}
H=\sqrt{\frac{2\,\alpha+1}{6}}\,k\,m\,\phi_i-\frac{2\,\alpha+1}{3(1+\alpha)}\,\frac{m^2}{k}\,t\,,
\end{equation}
besides, we are able to determine the following scale factor
\begin{equation}
a=a_i\,\exp\,\left\{\sqrt{\frac{2\,\alpha+1}{6}}\,k\,m\,\phi_i\,t-\frac{2\,\alpha+1}{6(1+\alpha)}\,\frac{m^2}{k}\,t^2\right\}\,.
\end{equation}

It is convenient to express time in terms of number of e-foldings $N$. We therefore obtain
\begin{equation}
N=\int_{\phi_i}^{\phi}\,\frac{H}{\dot{\phi}}\,d\,\phi=-\left(1+\alpha\right)\,k^2\,\int_{\phi_i}^{\phi}\frac{V}{V_\phi}\,d\phi\,,
\end{equation}
where we worked with both Eqs. from \eqref{18}. Once we are dealing with a Klein-Gordon potential, the last expression takes the form
\begin{equation}
N=\frac{k^2(1+\alpha)}{4}\,\left(\phi_i^2-\phi^2\right)\,,
\end{equation}
or
\begin{equation}
\phi=\sqrt{\phi_i^2-\frac{4\,N}{k^2\,(1+\alpha)}}\,.
\end{equation}
Such a relation is going to be very useful to compute the observable inflationary parameters. Firstly, using \eqref{18} - \eqref{18_2} into Eq. \eqref{11_1} we find
\begin{equation}
\epsilon_1=\frac{2}{\left(1+\alpha\right)\,k^2\,\phi^2}=\frac{2}{(1+\alpha)\, k^2 \phi_i^2-4\,N}\,,
\end{equation}
besides,
\begin{equation}
\epsilon_2=-2\epsilon_1=-\frac{4}{(1+\alpha)\, k^2 \phi_i^2-4\,N}\,.
\end{equation}

Thus, the observable inflationary parameters are given by
\begin{equation}
r=\frac{32}{(1+\alpha )\, k^2\, \phi_i^2-4\,N}\,;\qquad n_s=1-\frac{12}{(1+\alpha) k^2 \,\phi_i^2-4\,N}\,; \qquad n_T=-\frac{4}{(1+\alpha )\, k^2\, \phi_i^2-4\,N}\,,
\end{equation}
so, taking a desired $N=60$, we can use $\phi_i$ and $\alpha$ to determine several families of interesting parameters, as its shown in the table below.

\captionof{table}{Inflation observables for different values of $\phi_i$ and $\alpha$
}\label{tab:table2}
\begingroup
\setlength{\tabcolsep}{8pt} 
\renewcommand{\arraystretch}{1.5} 
\begin{tabular}{ |p{3cm}|p{3cm}|p{3cm}|p{3cm}| p{3cm}| }
 \hline

  $\phi_i\,k$  & $\alpha$   & $r$     & $n_{s}$ & $n_{T}$ \\
    \hline
     4 & 35.4 & 0.09346  & 0.96495  &-0.01168  \\
   
    8 & 8.11 &  0.09328 & 0.96501  &-0.01166  \\
    
    16 & 1.185 & 0.10020  & 0.96243  &-0.01252  \\
    32 & -0.431   & 0.09339 & 0.96498 & -0.01167\\  
    64 &-0.858    & 0.09367 & 0.96487  & -0.01171\\   
 \hline
 
\end{tabular}
\vspace{0.5cm}
\endgroup

As we can verify, the parameters presented in Table \eqref{tab:table2} are compatible with observable data from Planck \cite{planck}, as well as with our first attempt of an inflationary scenario for $f(R,T)$ (see Section $\ref{sec03}$).

\section{Final Remarks}
\label{sec06}
Inflation is the most well received theory describing the early Universe since it is successful in resolving many cosmological issues like the flatness problem, horizon problem, fine tuning problem, trans-Planckian problem, \textit{etc} \cite{guth,als,linde}. It is also successful in providing convincing answers to the formation of large scale structures form an almost linear to the presently observed non-linear regime \cite{mukh,guthpi,hawking,star}. Some of the predictions of inflation in the form of "inflation-observables" are also observationally verified and constrained. \\
A variety of theories capable of representing an inflationary scenario have been reported like the Starobinsky model, $\alpha$-attractors, Higgs field, \textit{etc}. Potentials of various types capable of producing an inflationary scenario have been exhaustively investigated and confronted with observational data \cite{hossain,martin1,martin2,geng,huang}.\\
Modified gravity theories like the Starobinsky model have been widely used to model an inflationary scenario. However, to the best of our knowledge, inflation has never been approached in $f(R,T)$ gravity. Hence, we tried to investigate the viability of the simplest minimal $f(R,T)$ gravity model of the form $f(R,T) = R + \eta T$ in addressing the inflationary scenario.\\
We found that the inflation observables are independent of the number of e-foldings $N$ but are very sensitive to the model parameter $\eta$ and requires extreme fine tuning to yield viable estimate of these observables. $\eta = -\kappa^{2}$ entails absence of any kind of inflation in the early Universe.\\
As a matter of completeness, we carefully analyzed the behavior of an inflation driven by a $f(R,T)$ coupled with a real scalar field. By taking the slow-roll approximation, we were able to generate interesting scenarios where a Klein Gordon potential leads to observable cosmological parameters. Such  parameters are compatible with the last set of data from Planck Collaboration \cite{planck}, and with our first approach. Therefore, our investigation unveils that in addition to the Ricci scalar and scalar fields, the trace of energy momentum tensor also play a major role in driving inflationary scenarios. Furthermore, the contribution of the trace of the energy-momentum tensor is also responsible to rescue an inflationary scenario for the Klein-Gordon potential, complementing the beautiful work of Ellis et al. \cite{ellis_2014} .  \\
As a possible extension of our work, one may be interested in modeling inflation with other $f(R,T)$ gravity models. Additionally, since the metric and Palatini versions of $f(R,T)$ gravity models are not equivalent, one may be interested in modeling inflationary scenarios with metric-affine $f(R,T)$ theories \cite{barry} as well. The present study can further be complemented by investigating inflationary cosmology in $f(R,T)$ gravity with the assistance from multi scalar fields \cite{kkvk} or from string-corrected axion dark matter \cite{sdvk}, which may generate interesting results and discussions.

\section*{ Acknowledgments} SB thanks Andrew R. Liddle for helpful discussions.  JRLS would like to thank CNPq for financial support, grant 420479/2018-0, and CAPES. PKS acknowledges CSIR, New Delhi, India for financial support to carry out the Research project [No.03(1454)/19/EMR-II Dt.02/08/2019]. PHRSM thanks CAPES for financial support. We are very much grateful to the honorable referee and the editor for the illuminating suggestions that have significantly improved our work in terms of research quality and presentation.

\end{document}